\documentclass[aps,pra,reprint,superscriptaddress,longbibliography]{revtex4-2}

\usepackage[utf8]{inputenc}	
\usepackage[T1]{fontenc}	
\usepackage{lmodern}
\usepackage{graphicx}
\usepackage{amsmath,amssymb}
\usepackage{setspace}
\usepackage{bm}             
\usepackage{booktabs} 		
\usepackage{nicefrac}		
\usepackage{microtype}		
\usepackage{lineno}		    
\usepackage{float}			
\usepackage[colorlinks=true,allcolors=blue]{hyperref}

\usepackage{scalerel}
\usepackage{tikz}
\usetikzlibrary{svg.path}

\definecolor{orcidlogocol}{HTML}{A6CE39}
\tikzset{
  orcidlogo/.pic={
    \fill[orcidlogocol] svg{M256,128c0,70.7-57.3,128-128,128C57.3,256,0,198.7,0,128C0,57.3,57.3,0,128,0C198.7,0,256,57.3,256,128z};
    \fill[white] svg{M86.3,186.2H70.9V79.1h15.4v48.4V186.2z}
                 svg{M108.9,79.1h41.6c39.6,0,57,28.3,57,53.6c0,27.5-21.5,53.6-56.8,53.6h-41.8V79.1z M124.3,172.4h24.5c34.9,0,42.9-26.5,42.9-39.7c0-21.5-13.7-39.7-43.7-39.7h-23.7V172.4z}
                 svg{M88.7,56.8c0,5.5-4.5,10.1-10.1,10.1c-5.6,0-10.1-4.6-10.1-10.1c0-5.6,4.5-10.1,10.1-10.1C84.2,46.7,88.7,51.3,88.7,56.8z};
  }
}

\newcommand\orcidicon[1]{\href{https://orcid.org/#1}{\mbox{\scalerel*{
\begin{tikzpicture}[yscale=-1,transform shape]
\pic{orcidlogo};
\end{tikzpicture}
}{|}}}}

\usepackage[overlay,absolute]{textpos}\newcommand\PlaceText[3]{\begin{textblock*}{10in}(#1,#2)#3\end{textblock*}}

\begin{document}
\PlaceText{12mm}{8mm}{\hspace{2.9cm} Phys. Rev. Applied \textbf{25}, 054029 (2026);
DOI: https://doi.org/10.1103/7zhl-5vv1}

\title{
Practical Countermeasure Against Attacks Exploiting Detection Efficiency Mismatch in Quantum Key Distribution
}

\author{Ben J. Taylor
\orcidicon{0009-0001-0231-7904}\ }
\email{ben.taylor@toshiba.eu}
\affiliation{Toshiba Europe Limited, 208 Cambridge Science Park, Cambridge CB4 0GZ, U.K.}
\affiliation{School of Physics, Engineering \& Technology and York Centre for Quantum Technologies, University of York, YO10 5FT York, U.K.}

\author{Peter R. Smith
\orcidicon{0000-0001-6007-2231}\ }
\affiliation{Toshiba Europe Limited, 208 Cambridge Science Park, Cambridge CB4 0GZ, U.K.}

\author{James F. Dynes}
\affiliation{Toshiba Europe Limited, 208 Cambridge Science Park, Cambridge CB4 0GZ, U.K.}

\author{Robert I. Woodward
\orcidicon{0000-0002-5026-494X}\ }
\affiliation{Toshiba Europe Limited, 208 Cambridge Science Park, Cambridge CB4 0GZ, U.K.}

\author{Marco Lucamarini
\orcidicon{0000-0002-7351-4622}\ }
\affiliation{School of Physics, Engineering \& Technology and York Centre for Quantum Technologies, University of York, YO10 5FT York, U.K.}

\author{R. Mark Stevenson}
\affiliation{Toshiba Europe Limited, 208 Cambridge Science Park, Cambridge CB4 0GZ, U.K.}

\author{Andrew J. Shields
\orcidicon{0000-0003-4100-3489}\ }
\affiliation{Toshiba Europe Limited, 208 Cambridge Science Park, Cambridge CB4 0GZ, U.K.}

\date{\today} 


\begin{abstract}
We demonstrate a practical countermeasure against a well-known class of attacks on quantum key distribution (QKD) systems that exploit detection efficiency 
mismatch, where the receiver's detectors do not exhibit identical responses to incoming photons across all degrees of freedom.
This class of quantum hacking strategies is broad and significantly includes the time-shift attack, which targets an arrival-time-dependent side channel at the receiver. The
four-state
countermeasure, previously only proven to be secure in theory, is implemented here on a GHz-clocked prototype QKD system and evaluated for its security and performance. We show that its presence enables almost complete recovery of the system's ideal secret key rate. Our results provide strong justification for adopting this countermeasure as a standard component in future scalable and practical QKD systems.
\end{abstract}

\maketitle


\section{Introduction}
 
Quantum key distribution (QKD) offers unconditionally secure communication guaranteed by the laws of quantum mechanics, providing the protocol's physical implementation does not deviate from its theoretical description.
As QKD technology has matured and progressed towards standardized real-world deployment \cite{dynes_cambridge_2019, sasaki_field_2011}, the field of implementation security, where security can be proven even with known hardware imperfections \cite{xu_secure_2020}, plays an increasingly vital role.

One approach to achieving this adapts security proofs to capture the impacts of imperfections, according to experimental characterisation. Alternatively, \textit{countermeasures} can be introduced in hardware or software, defined as modifications that close certain attack vectors or reduce side-channel information leakage to a potential eavesdropper (Eve). These countermeasures typically still require their own characterisations, e.g. bounding photon statistics of weak coherent pulses for implementing the decoy-state method \cite{lutkenhaus_quantum_2002, dynes_testing_2018}. 
Protocols such as measurement-device-independent \cite{lo_measurement-device-independent_2012, comandar_quantum_2016} or twin-field QKD \cite{lucamarini_overcoming_2018} grant adversaries full control over the detection system, but must deal with hardware imperfections at the transmitter.
Full device-independence in QKD remains extremely challenging in practice with current technology \cite{zapatero_advances_2023}.

In prepare-and-measure schemes such as BB84 \cite{bennett_quantum_2014}, 
the receiver is considered the most vulnerable element, as it necessarily accepts all light from the insecure quantum channel.
Consequently, it has been the primary target of quantum hacking attempts, such as blinding \cite{lydersen_thermal_2010, lydersen_hacking_2010, gao_ability_2022}, faked-state \cite{makarov_faked_2008, gerhardt_full-field_2011}, and Trojan horse \cite{gisin_trojan-horse_2006,sushchev_trojan-horse_2024} attacks.

The specific receiver imperfection considered in this paper is the mismatch in detection efficiency between two single-photon detectors, a side-channel commonly exploited in the quantum hacking literature, most notably and successfully via the time-shift attack \cite{zhao_quantum_2008, fung_security_2009} or in combination with a faked-state attack \cite{makarov_effects_2006}.

Here we investigate the performance and security of the best known countermeasure against these attacks, which is applicable to any active detection setup. The countermeasure has previously been addressed theoretically in \cite{fung_security_2009,lydersen_security_2010,makarov_effects_2006}, but despite reported usage in systems in \cite{nielsen_experimental_2001, wang_field_2010}, no implementation has been experimentally characterised against a detection efficiency mismatch-based attack, to the best of our knowledge. Our work therefore bridges an unresolved theory--experiment gap for high-speed QKD. This technique has been known by several names: \textit{detector symmetrization} \cite{lucamarini_implementation_2018}, \textit{randomization} \cite{ferreira_da_silva_safeguarding_2015}, or \textit{scrambling} \cite{ruhul_fatin_generalized_2021}, or \textit{four-state Bob} \cite{makarov_effects_2006, fung_security_2009}. The latter term is most descriptive of the hardware change made to the QKD receiver's phase modulator operation, so we adopt this terminology in our paper.

\subsection{Detection Efficiency Mismatch in QKD and the Time Shift Attack}

Original security proofs assume that Bob's detectors, typically avalanche photodiodes (APDs) in modern commercial systems, exhibit identical photo-detection responses across all degrees of freedom, including time, frequency and polarization. This assumption implies perfectly matched detection efficiencies, $\eta$, and dark-count rates, regardless of the properties of the incident light or the electronic settings applied to each APD for a given operating condition. In practice, achieving such stringent uniformity is extremely challenging.

\begin{figure}[t]
    \centering
    \includegraphics[width=\linewidth]{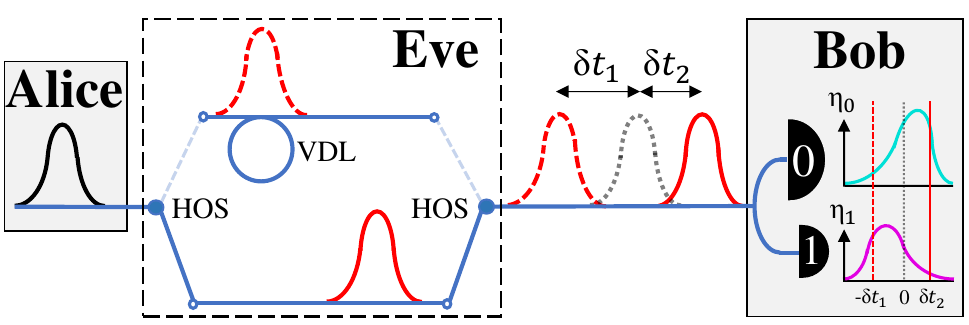}
    \caption{Conceptual schematic of a time-shift attack. HOS: high-speed optical switch; VDL: variable optical delay line. Eve shifts optical pulses generated by Alice in time by $+\delta t_1$ or $-\delta t_2$, chosen to maximise mismatch on Bob's two non-identical detectors, the efficiencies of which vary uniquely in time.
    \vspace{-0.3cm}
    }
    \label{fig0:concept}
\end{figure}

If this requirement is not met during a QKD system's operation, a bit-value-dependent (static) detection efficiency mismatch will be present. This leads to a disparity in the number of $1$s and $0$s that form the raw, and eventually the sifted, keys, with no influence from Eve \cite{makarov_preparing_2024}.
A significant deviation here would invalidate a critical cryptographic requirement of the final key's statistical randomness \cite{rusca_quantum_2024}.
In this static mismatch case, with only a known difference between the APD 0 and APD 1 efficiencies, $\eta_0$ and $\eta_1$, to consider, security is recovered simply by amending the key generation rate by a prefactor
\cite{fung_security_2009}
\vspace{-0.1cm}
\begin{equation}
    \label{eq:static_mismatch}
    \text{min} \left( \frac{\eta_0}{\eta_0+\eta_1}, \frac{\eta_1}{\eta_0+\eta_1} \right) .
\end{equation}

However, Eve is allowed to manipulate signals whilst in the quantum channel to affect the relative detection efficiency between Bob's APDs. By choosing an auxiliary degree of freedom to that of the encoding, she can gain significant knowledge of the key without increasing the quantum bit-error-rate (QBER) sufficiently to trigger Alice and Bob to abort their protocol.
This was successfully demonstrated in the time-shift attack (TSA) \cite{zhao_quantum_2008}, which is the most technologically-feasible strategy to exploit detection efficiency mismatch in QKD.

A conceptual schematic of this attack is shown in Fig. \ref{fig0:concept}.
Eve controls a high-speed optical switch inserted into the channel,
allowing her to choose between a shorter and longer optical path than the original fibre link, and so perturb the time-of-flight of each optical pulse passing through the channel.
Using tunable optical delay lines and low-loss fibre \cite{jain_attacks_2016}, Eve can optimize these two time-shifts such that they correspond to the worst case efficiency mismatches between between Bob's APDs.
These time-dependent mismatches may arise from different optical or electronic path lengths for signals going to each APD, or from inherent differences in the breakdown and threshold voltages of each semiconductor-based device \cite{hadfield_single-photon_2009}.
APDs are typically gated with an electronic driving signal each clock cycle, and each device will exhibit a unique time-dependent detection response across this gate.

When a significant mismatch is present, the probability that Bob detects a signal may depend as much on Eve's
time-shift choice as the agreement between Alice and Bob's basis choice. Whilst the key Alice and Bob output can still appear statistically random if Eve alternates time-shifts evenly, Eve will have greater knowledge than their privacy amplification has accounted for.

In the past two decades, several theoretical analyses have been presented that allow secure key rates in the presence of characterised detection mismatch \cite{fung_security_2009,trushechkin_security_2022,zhang_advances_2015,grasselli_quantum_2025};
a comprehensive summary of progress to date can be found in Table II of \cite{tupkary_phase_2025}.
A significant practical weakness with this proof-based approach is that every possible auxiliary dimension must be fully known and experimentally characterised for every given QKD system, before it can be considered secure. Additionally, it is highly likely that environmental conditions change over time, invalidating these characterisations. For this reason, practical countermeasures at a protocol level are preferable to calibration-based approaches where possible.

\subsection{Four-State Countermeasure}

In standard decoy-state BB84 QKD with time-bin/phase encoding, Alice uses an electro-optic modulator to encode a relative phase between a pair of optical pulses.
Alice's phases
\vspace{-0.2cm}
\begin{equation}
    \label{four_phases}
    \phi_A \in \left\{ 0, \frac{\pi}{2}, \pi, \frac{3\pi}{2} \right\},
\end{equation}
encode the states typically denoted $Z_0$, $X_0$, $Z_1$, and $X_1$. Bob then demodulates with 
\begin{equation}
    \label{two_phases}
    \phi_B \in \left\{ 0, \frac{\pi}{2} \right\},
\end{equation}
corresponding to a measurement in the $Z$ or $X$ basis respectively.
The sum $\phi_A + \phi_B$ determine the interference Bob will achieve at his interferometer output: constructive or destructive if the phases sum to 0 or $\pi$ respectively, (these contribute to sifted events), and random otherwise. This means that each \textit{bit-and-basis} choice from Alice has a deterministic detector allocation, given that Bob's \textit{basis} choice matched Alice's.

In the four-state countermeasure, the modulation values instead come from
the same set of four phases for $\phi_B$ as $\phi_A$ \cite{lagasse_secure_2005}.
The consequence of this is that now there are two possible combinations that will lead to a sifted bit for a given bit-and-basis choice at Alice.
As long as the random choice of bit value at Bob is recorded, he can map his output sifted bits from the \textit{physical} to \textit{logical} register in post-processing.
On average, any detection efficiency mismatch present between the physical detectors, regardless of the auxiliary degree of freedom responsible, should be erased in the logical detection events. As well as handling the static mismatch case described earlier, it should also shield the receiver against attacks from Eve that rely on exploiting the efficiency mismatch side-channel, for example with a faked-states attack 
\cite{makarov_faked_2008}.

The four-state countermeasure has been well-studied from a theoretical perspective,
and proven secure in \cite{fung_security_2009}. However, to the best of our knowledge, the security of this technique has not yet been evaluated on experimental results.
In the following, we replicate a time-shift attack on a GHz-clocked, fibre-based, phase-encoding QKD receiver, and compare the protocol's security in the cases that Bob demodulates with either two-states or four-states.

\begin{figure*}[t]
    \centering
    \includegraphics[width=0.9\textwidth]{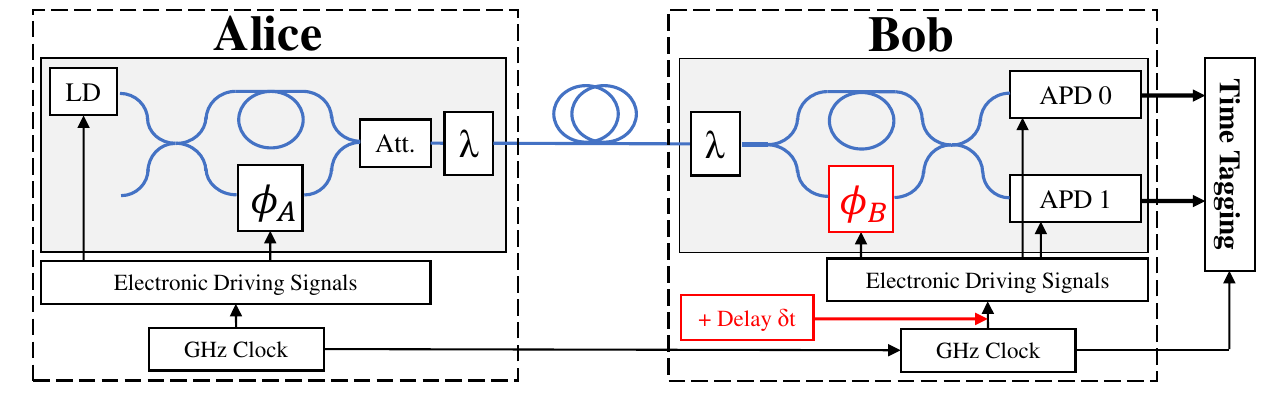}
    \caption{Key components for experimental characterisation of a fibre-based, GHz-clocked, phase-encoding QKD system's vulnerability to a time-shift attack. We sweep values of a relative delay, with resolution $\delta t$ = 4.5 ps, and add this to the global reference clock's signal, which is then supplied to all active optical components in Bob, whilst recording counts from the APD outputs.
    LD: laser diode; $\phi_A$, $\phi_B$: phase modulators; Att.: attenuation; $\lambda$: narrow-wavelength filter, here at 1550.12 $\pm$ 0.04 nm. 
    }
    \label{fig1:experimental_setup}
\end{figure*}

\section{Time-Shift Attack characterisation}

In Fig. \ref{fig1:experimental_setup}, we show key elements of the experimental setup we use to characterise the vulnerability of a fibre-based, GHz-clocked, QKD system to a TSA.
The optical configuration in Alice is standard for phase-encoding BB84 QKD, composed of a gain-switched laser diode producing phase-randomized optical pulses, into which relative phases are encoded within an asymmetric Mach-Zehnder interferometer (AMZI), before being attenuated to single photon intensity.

The receiver also contains standard optical components \cite{yuan_10-mbs_2018}. The phase modulator in Bob's AMZI can be programmed to use either two or four voltage levels.
After the AMZI, each beamsplitter output is connected to a single photon detector based on InGaAs 
APDs \cite{zhang_advances_2015, hadfield_single-photon_2009}.
All active modulation components in both Alice and Bob, as well as Bob's APDs, are
driven by local radiofrequency (RF) electronic driving signals. Alice and Bob are time synchronised with each other by use of a global reference clock signal.

By delaying the electronic driving signals that control the phase modulator and APD gating windows in Bob, we can replicate the effects of the TSA, as the pulse arrival time will be shifted with respect to the default modulation point. In our setup, we have separate control of the gating delays for Bob's phase modulator, APD 0, and APD 1, with a minimum resolution of 4.5 ps. We can therefore add or subtract delays with this resolution globally to all active components in the receiver.

As shown in Fig. \ref{fig1:experimental_setup}, we took the outputs of each logical APD and recorded clicks across the two channels of a time-to-digital converter (TDC) time tagger. In the case of the four-state countermeasure, the random bit-flip is undone in the post-processing hardware, which also handles enforced deadtimes and basis sifting. At each time-shift, we recorded 10Mb of total counts data.

\begin{figure}[h]
    \centering
    \includegraphics[width=1.02\linewidth]{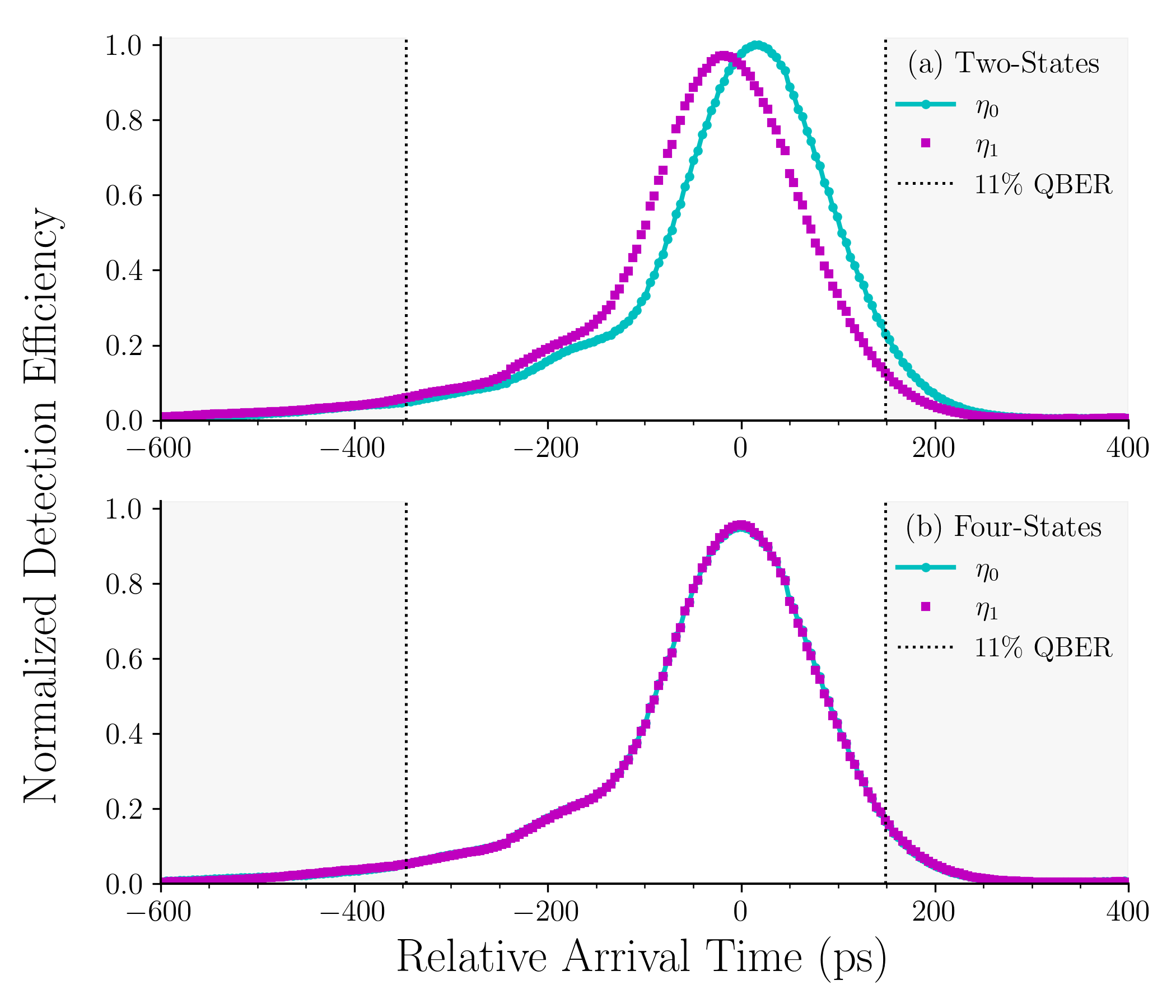}
    \vspace{-0.5cm}
    \caption{Normalized detection efficiency of APD 0 and APD 1 under time-shift relative to default, with steps of 4.5 ps across a range of 1 ns (full clock cycle), in both cases of (a) two-state and (b) four-state demodulation at Bob.
    We show bounds at which the QBER exceeded 11\% on average, such that the protocol would have aborted outside the white central region.
    Each data point comprises 10Mb of total counts data.}
    \vspace{-0.3cm}
    \label{fig2:efficiency_curves}
\end{figure}

Fig. \ref{fig2:efficiency_curves} illustrates the efficiency responses of the two detectors, APD 0 and APD 1, when the gating delays in Bob were swept in steps of 4.5 ps. Plots (a) and (b) correspond to
Bob demodulating using the standard two-states versus the countermeasure's four-states, respectively. 
To simulate the mismatch probed in the original TSA paper \cite{zhao_quantum_2008}, we deliberately displaced the individual APD gate delays away from their default positions and lowered the maximum efficiency of APD 1, thereby imposing a far more significant physical mismatch than would typically be present.
Despite this, the four-state countermeasure is extremely effective at statistically averaging the logical detection events, as evidenced by the efficiency curves for APD 0 and APD 1 in Fig. \ref{fig2:efficiency_curves}(b) now appearing virtually indistinguishable. Error bars were calculated from the square root of the count rates on each APD; these were too small to be visible on the plot and are therefore not shown.

\vspace{-0.2cm}
\subsection{Effect of Time-Shift Attack on Phase Modulator}

In high-speed QKD, a 1 GHz electronic driving signal will
have finite rise and fall times, and therefore will not be perfectly square-like.
As a consequence, far from the optimum gating position of Bob's phase modulator, all voltage levels will have lower amplitude than at their peak position, and hence the set of phases applied at Alice and Bob will differ. This will lower the interferometric visibility at Bob's APDs, an effect that becomes evident from an increased QBER at large time displacements, \textit{e.g.} in our experiment, the QBER reaches 50\% when the time-shift is approximately half the clock cycle, here 500ps.

\begin{figure}[!]
    \centering
    \includegraphics[width=0.9\linewidth]{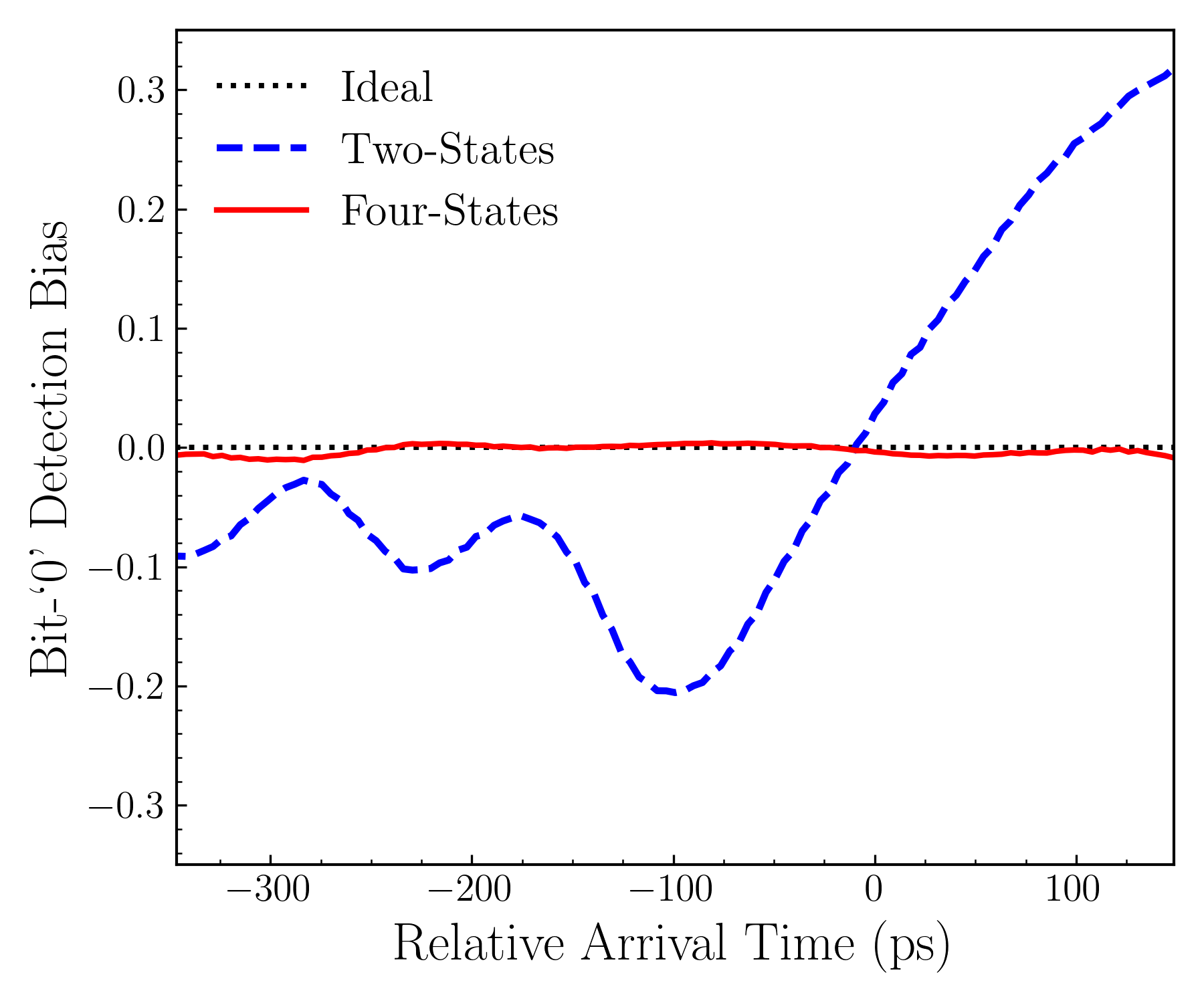}
    \vspace{-0.2cm}
    \caption{Bit-0 normalised detection bias across a 495 ps window, in increments of 4.5 ps.
    Whilst ideal APDs would exhibit no detection bias, the bias with two-states varies significantly, to a maximum of $\approx$ 30\%.
    This data subset corresponds to the central region in Fig. \ref{fig2:efficiency_curves}, where average QBER was below 11\%.}
    \label{fig3:bias}
    \vspace{-0.2cm}
\end{figure}

This effect allows us to identify bounds on the time displacements applied, where the average QBER exceeds 11\%,
shown in Fig. \ref{fig2:efficiency_curves}.
This is the most conservative cut-off where Alice and Bob can still distill a positive secret key \cite{shor_simple_2000}
(note that the tolerable QBER is closer to 7\% in real-world QKD with finite-size effects considered).

In Fig. \ref{fig3:bias} we show the bit-0 normalised detection bias with and without the countermeasure,
defined as the contrast $\frac{C_0 - C_1}{C_0 +C_1}$, with $C_0$, $C_1$, the count rates for each APD.
This data is a subset of that in Fig. \ref{fig2:efficiency_curves}, considering only the window inside the 11\% QBER bounds. The four-state setting follows the ideal bias lined very closely across this window, but the bias varies significantly in the two-state case and reaches a maximum of $\approx30$\%.
Again, error bars were too small to be visible.

\vspace{-0.2cm}
\section{Secret Key Rate Calculations}

The results from Fig. \ref{fig2:efficiency_curves} and Fig. \ref{fig3:bias} suggest the four-state countermeasure comprehensively erases any effects of physical detection efficiency mismatch in the sifted bits. To quantify the difference this would have on the secret key rates (SKR) for our QKD protocol, we follow the original proof outlined by Fung et al.in \cite{fung_security_2009} and built upon by Marcomini et al.\cite{marcomini_loss-tolerant_2025}, which utilizes the Procrustean Method, a filtering technique from entanglement distillation theory that attempts to orthogonalize non-orthogonal single-photon input states \cite{bennett_concentrating_1996}.

The inputs to this method are two diagonal matrices for each APD's efficiency responses across the set of all arrival times, with the basis defined by the time-shift resolution $\delta t$; we obtain the diagonal elements directly from the results shown in Fig. \ref{fig2:efficiency_curves}.
Following Marcomini et al.'s polarization-mismatch characterisation results  \cite{marcomini_loss-tolerant_2025}, where off-diagonal terms were roughly two orders of magnitude lower than the diagonal terms, we neglect off-diagonal contributions here.

\begin{figure}[b]
    \centering
    \includegraphics[width=\linewidth]{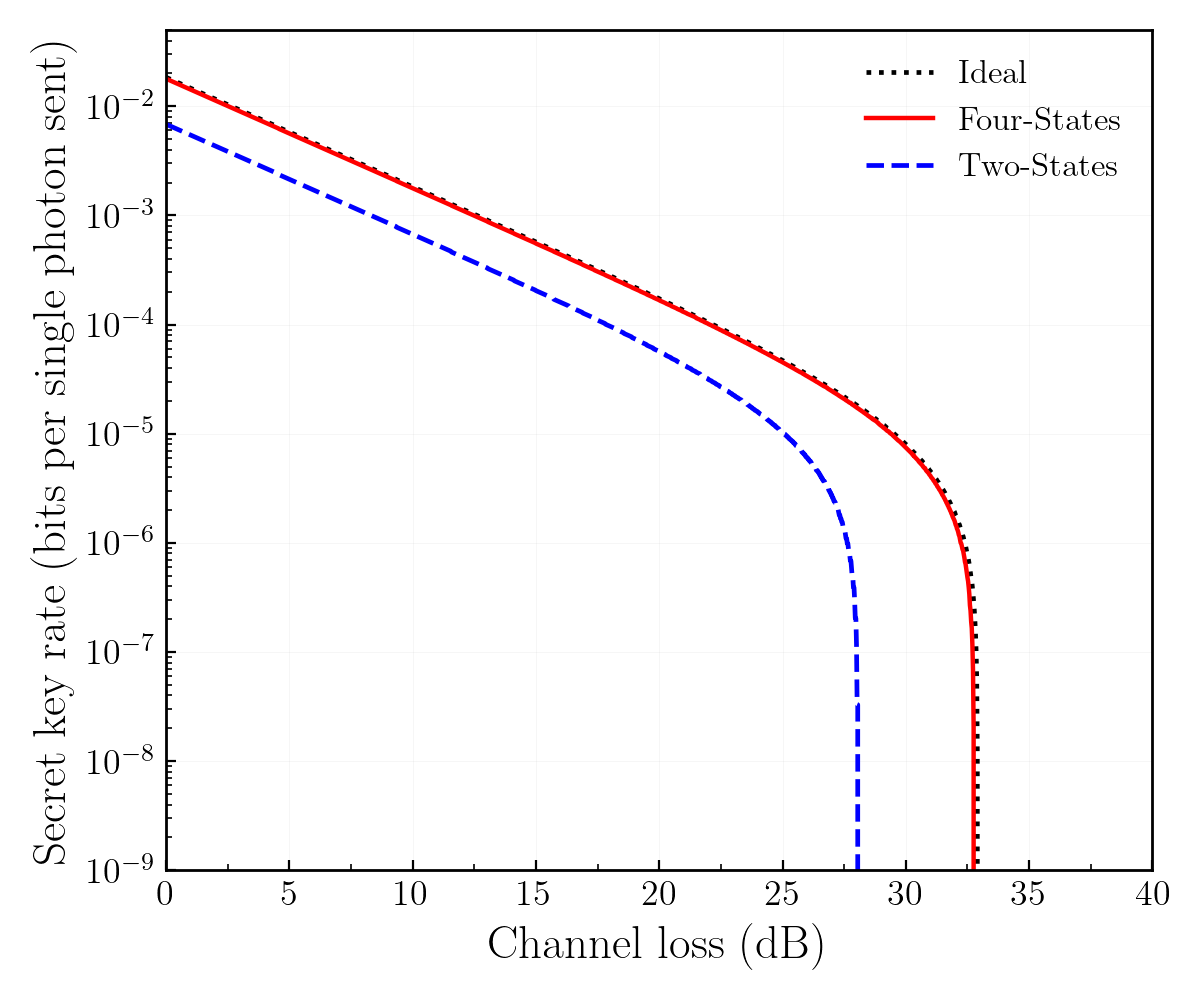}
    \caption{Asymptotic secret key rates (SKR) in bits per single photon sent against channel loss, accounting for detection efficiency mismatch. The four-state countermeasure recovers 97.1\% of the ideal case of no mismatch,
    whilst the rate is reduced to 38.3\% of ideal in the two-states case.}
    \label{fig:secret_key_rates}
\end{figure}

The method then numerically bounds the probability of successful Procrustean filtering and the increased phase error rate associated with this filter, respectively $p_{\mathrm{succ}}$ and $e_{\mathrm{phase}}$.
We set up two constrained optimisation problems using semi-definite programming (SDP) \cite{skrzypczyk_semidefinite_2023} to respectively minimize and maximise these values, subject to constraints that the phase and bit error rates due to Eve's actions must match those observed experimentally,  $e_{\mathrm{phase, obs}}$ and $e_{\mathrm{bit, obs}}$.
In the ideal case of no mismatch, we have $p_{succ}=1$ and $e_{\mathrm{phase}}$. 
Otherwise, $p_{\mathrm{succ}}$ and $e_{\mathrm{phase}}$ are respectively decreased and increased, in proportion to the significance of the efficiency mismatches across all time-shifts considered, again taken from the data subset shown displayed in Fig \ref{fig3:bias}.

The asymptotic SKR is computed, using values returned from the SDP, with
\begin{equation}
    R = \left[ p_\mathrm{succ}(1 - H_{2}(e_\mathrm{phase})) - f_{EC} H_2(e_\mathrm{bit}) \right],
    \label{eq:rate}
\end{equation}
where $H_{2}(x)$ is the binary entropy function, and $f_{EC}$ is the efficiency of classical error correction. 
In the ideal case, using QBERs of $e_{\mathrm{phase, obs}} = e_{bit,obs} = 0.03$ and $f_{EC}=1.10$ \cite{yuan_10-mbs_2018}, the maximum achievable rate is $R = 0.592$. The corresponding rates for the four-state and two-state cases are shown in Table \ref{tab:sdp_resolution_comparison}(b).
The four-state results recovers 97.1\% of the ideal SKR, whilst the rate is cut to 38.3\% of the ideal SKR in the two-states case.

In Fig. \ref{fig:secret_key_rates}, we compare how the asymptotic SKR performs with loss for these three scenarios. Here, the rate on detected single photons from Eq. (\ref{eq:rate}) is scaled by a prefactor $\eta$, encompassing channel transmissivity and total detection efficiency, with the observed error rates simulated as a function of loss, due to the contributions of dark counts.
Again, we see the four-state case very closely recovers the ideal SKR.

\begin{table}[H]
  \centering
  \begin{tabular}{ccc}
    \toprule
    & & \hspace{-3.0cm} (a) $\delta t$ = 49.5 ps \\
    \toprule
    \hspace{0.2cm} & \hspace{0.2cm} Two-states \hspace{0.2cm} & \hspace{0.2cm} Four-states \hspace{0.2cm} \\
    \toprule
    \textit{$p_{\mathrm{succ}}$} & 0.609 & 0.981 \\ 
    \textit{$e_{\mathrm{phase}}$} & 0.0475 & 0.0302 \\
    \textit{$R$} & 0.227 & 0.575 \\
    \bottomrule
  \end{tabular}
 
  \centering
  \begin{tabular}{ccc}
    & & \hspace{-3.0cm} (b) $\delta t$ = 4.5 ps \\
    \toprule
    \hspace{0.2cm} & \hspace{0.2cm} Two-states \hspace{0.2cm} & \hspace{0.2cm} Four-states \hspace{0.2cm} \\
    \toprule
    \textit{$p_{\mathrm{succ}}$} & 0.608 & 0.979 \\ 
    \textit{$e_{\mathrm{phase}}$} & 0.0470 & 0.0303 \\
    \textit{$R$} & 0.228 & 0.574 \\
    \bottomrule
  \end{tabular}
  \singlespacing
\caption{
Semi-definite program (SDP) results and asymptotic secret key rates ($R$ = secret bits per single photon received), showing the comparison between using time-shifts $\delta t$ of 49.5 or 4.5 picoseconds.
Here we use $e_{\mathrm{phase, obs}} = e_{bit,obs} = 0.03$ and $f_{EC}=1.10$, for which $R_{ideal}=0.592$.
}
\label{tab:sdp_resolution_comparison}
\end{table}

\vspace{-0.4cm}
\subsection{Minimum resolution of characterisation}

The proof by Fung et al. \cite{fung_security_2009} requires a narrow-wavelength filter is present at both the output of Alice and the input of Bob, when characterising the QKD system's vulnerability to the TSA.
This allows the Hilbert space to be treated as finite-dimensional, where in reality time is a continuous variable and a TSA is possible with arbitrary resolution.

We tested the accuracy of this assumption by using a higher-resolution sampling rate. The filter bandwidth in our setup was 10 GHz at 1550.12 nm, from which a minimum sampling resolution of 50 ps is derived, using the Nyquist-Shannon sampling theorem.
The experiment was performed using time-shifts of $\delta t$ = 4.5 ps, from which a set of subsampled shifts with $\delta t$ = 49.5 ps could be extracted. In Table \ref{tab:sdp_resolution_comparison}, we compare between the two resolutions, showing the results of the SDP and the SKR computed using Eq. (\ref{eq:rate}) in each case.
Approximately equivalent values were obtained, validating the finite-dimensional filter-based characterisation approach.

\section{Discussion}

We have demonstrated the practical security of the four-state countermeasure in a phase-encoding prototype QKD system. We have shown that the ideal SKR is almost completely recovered using this countermeasure despite the presence of severe physical detection efficiency mismatch, whilst the standard BB84 detector operation would see an extreme performance reduction. Our results provide strong justification for the adoption of this countermeasure as standard for future scalable and practical commercial, BB84-style, QKD systems.

Furthermore, our experimental verification of this technique's high performance is a substantial achievement, due to the high-speed nature of our implementation.
At GHz clock rates, it is not trivial to achieve the required phase modulation cleanly and with low residual error rates.
This distinguishes our demonstration from earlier work that reported usage of four-state demodulation with clock rates in the MHz range \cite{wang_field_2010}.
We note the technique may introduce a very slight increase in QBER, even under default QKD operation with the APD gating at their optimum timings alignments, likely due to the fact agreement is now required between Alice and Bob's phase modulator levels for two pairs of voltage levels rather than one.

As mentioned previously, security proofs now exist that account for detection efficiency mismatches with more realistic assumptions, including the incorporation of the decoy-state technique \cite{trushechkin_security_2022, lydersen_security_2010, zhang_security_2021, grasselli_quantum_2025}, and finite-size effects \cite{tupkary_phase_2025}. 
Future work could extend the analysis on results obtained from our experimental data, using more recent proof techniques.

However, whilst such proofs can explicitly quantify potential information leakage, we reiterate that for an implementation issue as broad and multi-faceted as detection efficiency mismatch, it is highly impractical to perform sophisticated characterisations, such as the one presented in this paper, for all possible auxiliary degrees of freedom (temporal, wavelength-based, and polarization-based mismatches, \textit{etc.} \cite{noauthor_bsi_2023}), for every single QKD system produced. Yet, this is what existing approaches require to guarantee secure operation, because each system will have a unique pair of single photon detectors.

On the other hand, the design-level robustness of the four-state countermeasure allows the same technique to be applied uniformly to all QKD systems, requiring one additional random number per clock cycle at the receiver in compensation for this significant practical advantage.
A further crucial difference to stress is that the principle of symmetrising logical detection events across the physical detection devices means that a QKD system can continue to securely generate key in the presence of severe mismatched detection efficiencies, even down to the case of using only a single detector in the receiver. 
The mismatch probed in our experiment was chosen to emulate the original TSA results \cite{zhao_quantum_2008}, but previous work shows eavesdropping strategies that exploit device calibration routines can induce extreme temporal mismatches \cite{jain_device_2011}, an attack vector against which only a QKD system using the four-state countermeasure would be resilient.
A characterisation-based proof approach, on the contrary, would not allow for any positive key to be established beyond a certain degree of mismatch.

A common criticism of this countermeasure is that it is vulnerable to a Trojan horse attack (THA) on the receiver \cite{makarov_effects_2006, fung_security_2009}, where Eve could read out Bob's phase modulator choices from the reflections of a bright light injection attack.
Yet, techniques to protect against a THA are well-known, for example using fibre delays \cite{dixon_quantum_2017}, optical isolation, wavelength filters, and watchdog detectors \cite{lucamarini_practical_2015}, and are required in a QKD system regardless of the modulation settings Bob uses.
Hence, we are unaware of any \textit{new} side-channel that is opened up with the four-state countermeasure.
We further stress that a practical time-shift attack is feasible with existing technology \cite{noauthor_bsi_2023}, and should correspondingly be considered the highest threat.
Closing the most accessible loopholes available to a quantum hacker, and thereby forcing them to resort to a more difficult and limited set of attacks, is significant progress in implementation security.

Finally, we caveat that the four-state countermeasure will not to be applicable to all QKD receivers and protocols. Previous analysis of free-space polarization-encoding QKD has shown that detection efficiency mismatch can still be present when all spatial modes are considered \cite{rau_spatial_2015}, a side-channel that does not exist in single-mode fibre.

\section*{Acknowledgements}
B.T. gratefully acknowledges funding from the Engineering and Physical Sciences Research Council. Toshiba Europe Limited. would like to acknowledge funding from Innovate UK's QAssure project (10102791).
M.L. would like to acknowledge the Engineering and Physical Sciences Research Council Integrated Quantum Networks Hub (EP/Z533208/1).
The authors thank Evan Lavelle for hardware support. B.T. thanks Joseph Dolphin, Martin Ward, and Matthew Winnel for helpful discussions.


%



\end{document}